# Phase transition of Na$_3$SbS$_4$ superionic conductor and its impact on ion transport

Dechao Zhang[1], Xiaoting Cao[1], Di Xu[1], Chuang Yu[2], Wentao Hu[1], Xinlin Yan[3], Jianli Mi[4], Bin Wen[1], Limin Wang[1], Long Zhang[1*]

[1]State Key Laboratory of Metastable Materials Science and Technology, Yanshan University, Qinhuangdao, Hebei 066004, China

[2]Department of Radiation Science and Technology, Delft University of Technology, Mekelweg 15, Delft 2629 JB, The Netherlands

[3]Institute of Solid State Physics, Vienna University of Technology, Wiedner Hauptstr. 8-10, 1040 Vienna, Austria

[4]Institute for Advanced Materials, School of Materials Science and Engineering, Jiangsu University, Zhenjiang, Jiangsu 212013, China

[*] Corresponding author. Tel: +86 335 8057047. Email: lzhang@ysu.edu.cn

**Abstract**

In this work, a comprehensive research coupling experiment and computation has been performed to understand the phase transition of Na$_3$SbS$_4$ and to synthesize cubic Na$_3$SbS$_4$ (*c*-Na$_3$SbS$_4$), a high temperature phase of Na$_3$SbS$_4$ that is difficult to be preserved from high temperature. The formation of *c*-Na$_3$SbS$_4$ is verified by Rietveld refinement and electrochemical impedance spectroscopy. Both experiment and theoretical calculation reveal that the ionic conductivity of *c*-Na$_3$SbS$_4$ is higher than that of *t*-Na$_3$SbS$_4$, though the values are in the same order of magnitude. Both structures allow fast ion transport.

**Introduction**

The earth abundant sodium as well as its proper redox potential have stimulated the development of sodium ion batteries for high capacity rechargeable batteries used in large-scale energy storage, including electric vehicles and electrical grid energy storage.[1-4] In parallel, further development toward high energy density and large-scale batteries has raised safety concerns regarding conventional rechargeable batteries with liquid electrolytes because of the instability of organic liquid electrolytes and the hazards of electrolyte leakage.[5-11] All-solid-state batteries have attracted worldwide attention because of several advantages, including absence of leakage, wide electrochemical windows, high thermal stability, as well as direct-stacking of batteries for high operating voltages and volumetric miniaturization.[12-15] Therefore, the merits of all-solid-state sodium ion batteries (ASSIBs) are ideal for high capacity large-scale energy storage and severe safety-concerned application, e.g., public transport. One of the main challenges in the development of ASSIBs is the lack of a sodium solid electrolyte with high ionic conductivity at room temperature.[16, 17] It is essential to develop new solid electrolyte with high ionic conductivity through carful composition and structure design.

Among various fast ion conductors, sulfides have been studied intensively as one of the promising inorganic solid electrolytes.[14, 15, 18-21] This type of materials takes the advantages of high ionic conductivity, low sintering temperature, and negligible grain-boundary resistance that is compacted by conventional cold-press conditions. Such room temperature pressing technique is favorable for assembling all-solid-state batteries. However, up to now only a few sulfide-based or sulfide-analogous Na-solid electrolytes have been developed, e.g. $Na_3PS_4$ and its related compounds,[22-34] Se-based compounds,[35-38] $Na_3SbS_4$,[39-41] $Na_{10}SnP_2S_{12}$,[42] and $Na_{10}GeP_2S_{12}$.[43]

Cubic Na$_3$PS$_4$,[22] synthesized by Prof. Hayashi and co-workers in 2012, is regarded as the milestone of sulfide sodium superionic conductors with high ionic conductivity at room temperature (0.2 mS/cm). Further improvements with the values of 0.46[24] and 0.74[44] mS/cm were achieved in the same group by using high purity starting materials and doping Si, respectively. These values are two orders of magnitude higher than that of tetragonal phase Na$_3$PS$_4$.[45] Another cubic compound Na$_3$PSe$_4$ exhibits even higher ionic conductivity with the value over 1 mS/cm.[37, 38] Although these results suggest that cubic structure is more promising for faster ion conduction in contrast to tetragonal structure, the subsequent computational investigations indicate that cubic and tetragonal Na$_3$PS$_4$ show comparable ionic conductivity.[29, 31] On the other hand, defects are thought to play a more important role than phase structure in ion diffusion mechanism.[29-32, 37, 46, 47] However, the impact of phase structure on ion diffusion have not yet been directly clarified by means of experimental comparison.

Very recently, tetragonal Na$_3$SbS$_4$ ($t$-Na$_3$SbS$_4$) has been investigated in our and some other groups and found to be a very promising sodium solid electrolyte.[39-41] The structural analysis shows a space group $P$-42$_1c$ with the lattice parameters $a,b$ = 7.1597 (5) Å and $c$ = 7.2906(6) Å.[39] This dimension is just slightly elongated by $c$ axis from its cubic symmetry.[48] We were thus motivated to systematically investigate the phase transition of Na$_3$SbS$_4$ between the cubic and the tetragonal structures and compare their ion transport properties. Here, we report synthesis of both $t$- and $c$-Na$_3$SbS$_4$ and a comprehensive study on phase transition in Na$_3$SbS$_4$ through differential scanning calorimetry (DSC), X-ray powder Diffraction (XRD), as well as electrochemical impedance spectroscopy (EIS). The influence factors on ion transport are studied including the phase structure and the crystallization status. In addition, theoretical simulations were performed to further understand the Na$^+$ diffusion performance in $c$- and $t$-Na$_3$SbS$_4$.

**Methods**

**Computational methods**

First principles calculations based on density functional theory (DFT) were performed by employing the Vienna ab initio simulation package (VASP). The exchange-correlation energy was treated within the generalized gradient approximation (GGA), using the functional of Perdew-Burke-Ernzerhof (PBE). The kinetic energy cutoff is 280 eV. Structures were optimized with the conjugate gradient method. The energy convergence value between two consecutive steps was set as $10^{-6}$ eV. A maximum force of $10^{-2}$ eV/Å was allowed on each atom. The Brillouin zone was sampled using a 7×7×7 Monkhorst-Pack $k$-point scheme for relaxing a single unit cell, and a 2×2×2 Monkhorst-Pack $k$-point scheme for a 2×2×2 supercell. For AIMD calculations, a 2 × 2 × 2 supercell was used with a k-point mesh of 1 × 1 × 1. The time between AIMD steps was 2 fs, and the total simulation time was 70 ps. The AIMD simulations were performed in the NVT ensemble with velocity scaling at every time step. In all cases an equilibration time of 2.5 ps was applied to allow the system to reach equilibrium during the AIMD simulations.

**Synthesis of $Na_3SbS_4$**

Tetragonal $Na_3SbS_4$ precursor was synthesized by solid state reaction.[39] Na granule (AR, Sinopharm), Sb powder (99.999%, Sinopharm), and S powder (99.999%, Alfa) were used as starting materials. These materials were weighed in an argon-filled glove box with $H_2O$, $O_2$ < 0.5 ppm according to the molar ratio of Na:Sb:S = 3:1:4 and loaded into a glassy carbon crucible vacuum-sealed in a quartz tube. The mixture was slowly heated to 700 °C, dwelled for 12 h, and cooled down naturally in the furnace. The resulted sample is designated as MNC1 and used as the precursors to prepare $c$-$Na_3SbS_4$.

**Materials characterization**

X-ray diffraction (XRD) measurements were carried out with a Rigaku D/MAX-2500/PC (Cu K$_a$, 40 kV 200mA). *In situ* XRD were measured from room temperature to 250 °C with 30 min dwelling before recording. Rietveld refinements were performed using the FULLPROF program to determine the filling fraction (FF) and lattice parameter. Raman scattering measurements were performed using a Renishaw inVia system with a 514.5 nm excitation source. DSC profiles were recorded on a Perkin-Elmer DSC8000 with a scan rate of 10 °C/min.

**Electrochemical measurement**

AC impedance was determined using an impedance analyzer (Princeton P4000) in the frequency range of 0.1 Hz – 2 MHz. Temperature dependence of conductivity was recorded during the heating scan in the range of room temperature to 215 °C. The pellets for measurements were cold-pressed at ~300 MPa with stainless-steel foils placed on both sides of the pellets. Stainless-steel rods were used as current collectors.

**Results and Discussion**

DSC measurements of MNC1 show that a reversible phase transition appears at 170 °C during heating and 169 °C during cooling, respectively, though the peak intensity is weak. These phase transitions relate to the tetragonal phase transformed to the cubic phase and the cubic phase inversely transformed to the tetragonal phase, respectively. $c$-Na$_3$SbS$_4$ is a high temperature phase of $t$-Na$_3$SbS$_4$.

Therefore, we have tried to prepare $c$-Na$_3$SbS$_4$ including high energy ball milling (HBM) and low energy ball milling (LBM). The sample subjected to HBM with a rotation speed of 250 rpm is designated as HBM1. Figure 1 displays the XRD profile of HBM1. Fitting the pattern using Rietveld refinement, cubic model is more accurate than tetragonal model. The lattice parameters of $c$-Na$_3$SbS$_4$ are $a = b = c = 0.71910$ (7) nm, which are close to those of $t$-Na$_3$SbS$_4$.[39] Each unit cell consists of two SbS$_4$ units with Sb atoms situated on the 2$a$ sites and S atoms on the 8$c$ sites, and Na atoms occupying the 6$b$ sites. The diffusion channels formed by SbS$_4$ tetrahedra provide orthogonal 3D conduction pathways for Na$^+$.

LBM was performed with a slow rotation speed of 100 rpm to reduce the grain size as close as possible to that of HBM1 meanwhile keeps a tetragonal-dominated structure. Figure 2 shows the XRD profiles of the sample prepared by LBM for 20 h. The sample is designated as LBM1. The precursor (MNC1) and HBM1 were presented for comparison. The peak widths were broadened after LBM in contrast to those of the precursor. The crystallite size was calculated to be 18 nm from 'SYSSIZE'.[49] This value is comparable to that of HBM1 (16 nm).

The ion transprot property of $c$-Na$_3$SbS$_4$ (HBM1) and $t$-Na$_3$SbS$_4$ (LBM1) are illustrated in Figure 3. Room temperature Nyquist impedance plots (Figure 3a) contains a small semicircle in the high-frequency region and a linear spike in the low-frequency region, associated with a small grain boundary resistance and capacitive behavior similar to the blocking electrodes (a typical ionic conductor). The total resistance (including the bulk and grain-boundary resistances) of $c$- and $t$-Na$_3$SbS$_4$ are 80 Ω and 127 Ω, respectively, from which the ionic conductivities are calculated to be 2.8 and 1.77 mS/cm. The inset of Figure 3a shows the ionic conductivity ($\sigma$) of $t$-Na$_3$SbS$_4$ as a function of LBM time. The ionic conductivity increases monotonously with increasing milling time from 0 to 20 h, indicating a benefit of reduction of grain size.

Figure 3b shows the temperature dependence of the ionic conductivity of $c$- and $t$-Na$_3$SbS$_4$. The former demonstrates a linear dependence of log$\sigma$ vs. (1/$T$) following the Arrhenius law, while the latter displays a distinct inflection point at about 100 °C generating two linear sections. The turning point we interpret as a result of the appearance of phase transition, which leads to a narrowed difference of ionic conductivity between the two samples at high temperature. The phase transition temperature of $t$-Na$_3$SbS$_4$ transformed from tetragonal to cubic measured by AC impedance method is lower than that measured by DSC. This may be induced by the pressure loaded during AC impedance measurements. The activation energy for $c$-Na$_3$SbS$_4$ is $E_a$ = 0.06 eV and for $t$-Na$_3$SbS$_4$ is $E_a$ = 0.25 eV for tetragonal segment and $E_a$ = 0.08 eV for cubic segment. The values of the cubic phase are close to each other for the two samples. In addition, samples were annealed at 120 °C to evaluate the influence of the glasses on ion transport, as shown in the inset of Figure 3b. MNC1 is glass-free since no glass transition peak was observed. The ionic conductivity of MNC1 before and after annealing keeps almost unchanged, but that of HBM1 increases after annealing. The enhanced conductivity is attributed to the precipitation of crystalline phase from glasses by heat treatment, which has proven to be a useful way to enhance ion transport.[12]

AIMD simulations were also performed to study Na$^+$ diffusivity in Na$_3$SbS$_4$, as illustrated in Figure 3c. Our preliminary work reveals that the diffusivity is negligible in stoichiometric Na$_3$SbS$_4$ with perfect structure, while it is significantly enhanced in vacancy contained structure. This is analogous with those of Na$_3$PS$_4$ and Na$_3$PSe$_4$ previously reported.[29, 31, 37] The diffusion ability of $c$- and $t$-Na$_{2.94}$SbS$_4$, with the assumption of no phase transition, is therefore modeled with the structure containing 2% Na vacancies. The vacancies were set to occupy in 6$b$ and 2$a$ sites for $c$- and $t$-Na$_{2.94}$SbS$_4$, respectively. Na$^+$ diffusion coefficient D was derived from the mean-squared displacement (MSD) data shown in the inset of Figure 3c. Na$^+$ diffusion increases with increasing temperature for both structures. Na$^+$ diffusivity in cubic structure is higher than that in tetragonal structure when T < 800 K. After that, the tetragonal structure shows slightly higher diffusivity. According to the results in Figure 3b, supposing no phase transition appears for $t$-Na$_3$SbS$_4$ at high temperature, i.e., extrapolating the plot of log$\sigma$ versus (1/T) at lower temperature section to a high temperature region, the ionic conductivity of $t$-Na$_3$SbS$_4$ would higher than that of $c$-Na$_3$SbS$_4$.

## 4. Conclusions

To summarize, a detail study has been carried out on synthesis, phase transformation, and ion transport for $c$- and $t$-Na$_3$SbS$_4$. DSC and *in situ* XRD demonstrate that the structural transition is reversible between cubic and tetragonal. $c$-Na$_3$SbS$_4$ is a high temperature phase of $t$-Na$_3$SbS$_4$. The evidence of formation of cubic phase has been addressed by Rietveld refinement of XRD profile fitted with both cubic and tetragonal models, where the former runs obvious lower reliability factors than the latter. Due to the phase transition from tetragonal to cubic in $t$-Na$_3$SbS$_4$ at high temperature, temperature dependence of the ionic conductivity for $c$-Na$_3$SbS$_4$ follows the Arrhenius law while that for $t$-Na$_3$SbS$_4$ appears two line sections with a distinct turning point. Both calculating and experimental data demonstrate that the ionic conductivity of the cubic phase is higher than that of the tetragonal phase, though the values are in the same order of magnitude.

**Acknowledgments**

This work was supported by the Science Foundation of Hebei Education Department (ZD2016033)..

**References**

1. Larcher, D.; Tarascon, J. M. Towards greener and more sustainable batteries for electrical energy storage. *Nat. Chem.* **2015**, *7*, 19–29.

2. Kundu, D.; Talaie, E.; Duffort, V.; Nazar, L. F. The emerging chemistry of sodium ion batteries for electrochemical energy storage. *Angew. Chem. Int. Ed. Engl.* **2015**, *54*, 3431–48.

3. Wen, Z.; Hu, Y.; Wu, X.; Han, J.; Gu, Z. Main Challenges for High Performance NAS Battery: Materials and Interfaces. *Adv. Funct. Mater.* **2013**, *23*, 1005–1018.

4. Pan, H.; Hu, Y.-S.; Chen, L. Room-temperature stationary sodium-ion batteries for large-scale electric energy storage. *Energy Environ. Sci.* **2013**, *6*, 2338–2360.

5. Bachman, J. C.; Muy, S.; Grimaud, A.; Chang, H.-H.; Pour, N.; Lux, S. F.; Paschos, O.; Maglia, F.; Lupart, S.; Lamp, P.; Giordano, L.; Shao-Horn, Y. Inorganic Solid-State Electrolytes for Lithium Batteries: Mechanisms and Properties Governing Ion Conduction. *Chem. Rev.* **2016**, *116*, 140–162.

6. Kim, J. G.; Son, B.; Mukherjee, S.; Schuppert, N.; Bates, A.; Kwon, O.; Choi, M. J.; Chung, H. Y.; Park, S. A review of lithium and non-lithium based solid state batteries. *J. Power Sources* **2015**, *282*, 299–322.

7. Kalhoff, J.; Eshetu, G. G.; Bresser, D.; Passerini, S. Safer Electrolytes for Lithium-Ion Batteries: State of the Art and Perspectives. *Chemsuschem* **2015**, *8*, 2154–2175.

8. Jung, Y. S.; Oh, D. Y.; Nam, Y. J.; Park, K. H. Issues and Challenges for Bulk-Type All-Solid-State Rechargeable Lithium Batteries using Sulfide Solid Electrolytes. *Isr. J. Chem.* **2015**, *55*, 472–485.

9. Zhang L.; Yang K.; Dong J.; Lu L. Recent developments in thio-LISICON solid electrolytes. *J. Yanshan Univ.* **2015**, *39*, 95–106.

10. Yao, X.; Huang, B.; Yin, J.; Peng, G.; Huang, Z.; Gao, C.; Liu, D.; Xu, X. All-solid-state lithium batteries with inorganic solid electrolytes: Review of fundamental science. *Chin. Phys. B.* **2016**, *25*, 018802.

11. Vignarooban, K.; Kushagra, R.; Elango, A.; Badami, P.; Mellander, B. E.; Xu, X.; Tucker, T. G.; Nam, C.; Kannan, A. M. Current trends and future challenges of electrolytes for sodium-ion batteries. *Int. J. Hydrogen Energy* **2016**, *41*, 2829–2846.

12. Tatsumisago, M.; Nagao, M.; Hayashi, A. Recent development of sulfide solid electrolytes and interfacial modification for all-solid-state rechargeable lithium batteries. *J. Asian Ceram. Soc.* **2013**, *1*, 17–25.

13. Takada, K. Progress and prospective of solid-state lithium batteries. *Acta Mater.* **2013**, *61*, 759–770.

14. Hayashi, A.; Sakuda, A.; Tatsumisago, M. Development of Sulfide Solid Electrolytes and Interface Formation Processes for Bulk-Type All-Solid-State Li and Na Batteries. *Front. Energy Res.* **2016**, *4*, 25.

15. Lin, Z.; Liang, C. Lithium-sulfur batteries: from liquid to solid cells. *J. Mater. Chem. A.* **2015**, *3*, 936–958.

16. Slater, M. D.; Kim, D.; Lee, E.; Johnson, C. S. Sodium-Ion Batteries. *Adv. Funct. Mater.* **2013**, *23*, 947–958.

17. Palomares, V.; Casas-Cabanas, M.; Castillo-Martínez, E.; Han, M. H.; Rojo, T. Update on Na-based battery materials. A growing research path. *Energy Environ. Sci.* **2013**, *6*, 2312.

18. Zhang, S.; Ueno, K.; Dokko, K.; Watanabe, M. Recent Advances in Electrolytes for Lithium-Sulfur Batteries. *Adv. Energy Mater.* **2015**, *5*, 1500117.

19. Kamaya, N.; Homma, K.; Yamakawa, Y.; Hirayama, M.; Kanno, R.; Yonemura, M.; Kamiyama, T.; Kato, Y.; Hama, S.; Kawamoto, K.; Mitsui, A. A lithium superionic conductor. *Nat. Mater.* **2011**, *10*, 682–686.

20. Tatsumisago, M.; Hayashi, A. Sulfide Glass-Ceramic Electrolytes for All-Solid-State Lithium and Sodium Batteries. *Int. J. Appl. Glass Sci.* **2014**, *5*, 226–235.

21. Wang, Y.; Richards, W. D.; Ong, S. P.; Miara, L. J.; Kim, J. C.; Mo, Y.; Ceder, G. Design principles for solid-state lithium superionic conductors. *Nat. Mater* **2015**, *14*, 1026–1031.

22. Hayashi, A.; Noi, K.; Sakuda, A.; Tatsumisago, M. Superionic glass-ceramic electrolytes for room-temperature rechargeable sodium batteries. *Nat. Commun.* **2012**, *3*, 856.

23. Berbano, S. S.; Seo, I.; Bischoff, C. M.; Schuller, K. E.; Martin, S. W. Formation and structure of $Na_2S+P_2S_5$ amorphous materials prepared by melt-quenching and mechanical milling. *J. Non-Cryst. Solids.* **2012**, *358*, 93–98.

24. Hayashi, A.; Noi, K.; Tanibata, N.; Nagao, M.; Tatsumisago, M. High sodium ion conductivity of glass–ceramic electrolytes with cubic $Na_3PS_4$. *J. Power Sources.* **2014**, *258*, 420–423.

25. Jha, P. K.; Pandey, O. P.; Singh, K. Crystallization and Glass Transition Kinetics of $Na_2S$-$P_2S_5$-Based Super-Ionic Glasses. *Part. Sci. Technol.* **2014**, *33*, 166–171.

26. Noi, K.; Hayashi, A.; Tatsumisago, M. Structure and properties of the $Na_2S$–$P_2S_5$ glasses and glass–ceramics prepared by mechanical milling. *J. Power Sources.* **2014**, *269*, 260–265.

27. Tanibata, N.; Noi, K.; Hayashi, A.; Kitamura, N.; Idemoto, Y.; Tatsumisago, M. X-ray Crystal Structure Analysis of Sodium-Ion Conductivity in $94\,Na_3PS_4 \cdot 6\,Na_4SiS_4$ Glass-Ceramic Electrolytes. *ChemElectroChem.* **2014**, *1*, 1130–1132.

28. Hibi, Y.; Tanibata, N.; Hayashi, A.; Tatsumisago, M. Preparation of sodium ion conducting $Na_3PS_4$–NaI glasses by a mechanochemical technique. *Solid State Ionics.* **2015**, *270*, 6–9.

29. Zhu, Z.; Chu, I.-H.; Deng, Z.; Ong, S. P. Role of Na+Interstitials and Dopants in Enhancing the Na+Conductivity of the Cubic Na$_3$PS$_4$ Superionic Conductor. *Chem. Mater.* **2015**, *27*, 8318–8325.

30. Chu, I.-H.; Kompella, C. S.; Nguyen, H.; Zhu, Z.; Hy, S.; Deng, Z.; Meng, Y. S.; Ong, S. P. Room-Temperature All-solid-state Rechargeable Sodium-ion Batteries with a Cl-doped Na$_3$PS$_4$ Superionic Conductor. *Sci. Rep.* **2016**, *6*, 33733.

31. de Klerk, N. J. J.; Wagemaker, M. Diffusion Mechanism of the Sodium-Ion Solid Electrolyte Na$_3$PS$_4$ and Potential Improvements of Halogen Doping. *Chem. Mater.* **2016**, *28*, 3122–3130.

32. Yu, C.; Ganapathy, S.; de Klerk, N. J. J.; van Eck, E. R. H.; Wagemaker, M. Na-ion dynamics in tetragonal and cubic Na$_3$PS$_4$, a Na-ion conductor for solid state Na-ion batteries. *J. Mater. Chem. A.* **2016**, *4*, 15095–15105.

33. Wenzel, S.; Leichtweiss, T.; Weber, D. A.; Sann, J.; Zeier, W. G.; Janek, J. Interfacial Reactivity Benchmarking of the Sodium Ion Conductors Na$_3$PS$_4$ and Sodium beta-Alumina for Protected Sodium Metal Anodes and Sodium All-Solid-State Batteries. *ACS Appl. Mater. Interfaces.* **2016**, *8*, 28216–28224.

34. Bo, S.-H.; Wang, Y.; Ceder, G. Structural and Na-ion conduction characteristics of Na$_3$PS$_x$Se$_{4-x}$. *J. Mater. Chem. A.* **2016**, *4*, 9044–9053.

35. Pompe, C.; Pfitzner, A. Na$_3$SbSe$_3$: Synthesis, Crystal Structure Determination, Raman Spectroscopy, and Ionic Conductivity. *Z. Anorg. Allg. Chem.* **2012**, *638*, 2158–2162.

36. Kim, S. K.; Mao, A.; Sen, S.; Kim, S. Fast Na-Ion Conduction in a Chalcogenide Glass–Ceramic in the Ternary System Na$_2$Se–Ga$_2$Se$_3$–GeSe$_2$. *Chem. Mater.* **2014**, *26*, 5695–5699.

37. Bo, S.-H.; Wang, Y.; Kim, J. C.; Richards, W. D.; Ceder, G. Computational and Experimental Investigations of Na-Ion Conduction in Cubic Na$_3$PSe$_4$. *Chem. Mater.* **2016**, *28*, 252–258.

38. Zhang, L.; Yang, K.; Mi, J.; Lu, L.; Zhao, L.; Wang, L.; Li, Y.; Zeng, H. Na$_3$PSe$_4$: A Novel Chalcogenide Solid Electrolyte with High Ionic Conductivity. *Adv. Energy Mater.* **2015**, *5*.

39. Zhang, L.; Zhang, D.; Yang, K.; Yan, X.; Wang, L.; Mi, J.; Xu, B.; Li, Y. Vacancy-Contained Tetragonal Na$_3$SbS$_4$ Superionic Conductor. *Adv. Sci.* **2016**, *3*, 1600089.

40. Banerjee, A.; Park, K. H.; Heo, J. W.; Nam, Y. J.; Moon, C. K.; Oh, S. M.; Hong, S. T.; Jung, Y. S. Na$_3$SbS$_4$: A Solution Processable Sodium Superionic Conductor for All-Solid-State Sodium-Ion Batteries. *Angew. Chem. Int. Ed.* **2016**, *55*, 9634–8.

41. Wang, H.; Chen, Y.; Hood, Z. D.; Sahu, G.; Pandian, A. S.; Keum, J. K.; An, K.; Liang, C. An Air-Stable Na$_3$SbS$_4$ Superionic Conductor Prepared by a Rapid and Economic Synthetic Procedure. *Angew. Chem. Int. Ed.* **2016**, *55*, 8551–8555.

42. Richards, W. D.; Tsujimura, T.; Miara, L. J.; Wang, Y.; Kim, J. C.; Ong, S. P.; Uechi, I.; Suzuki, N.; Ceder, G. Design and synthesis of the superionic conductor Na$_{10}$SnP$_2$S$_{12}$. *Nat. Commun.* **2016**, *7*, 11009.

43. Kandagal, V. S.; Bharadwaj, M. D.; Waghmare, U. V. Theoretical prediction of a highly conducting solid electrolyte for sodium batteries: Na$_{10}$GeP$_2$S$_{12}$. *J. Mater. Chem. A* **2015**, *3*, 12992–12999.

44. Tanibata, N.; Noi, K.; Hayashi, A.; Tatsumisago, M. Preparation and characterization of highly sodium ion conducting Na$_3$PS$_4$–Na$_4$SiS$_4$ solid electrolytes. *RSC Adv.* **2014**, *4*, 17120.

45. Jansen, M.; Henseler, U. Synthesis, structure determination, and ionic conductivity of sodium tetrathiophosphate. *J. Solid State Chem.* **1992**, *99*, 110–119.

46. Holzmann, T.; Schoop, L. M.; Ali, M. N.; Moudrakovski, I.; Gregori, G.; Maier, J.; Cava, R. J.; Lotsch, B. V. Li$_{0.6}$[Li$_{0.2}$Sn$_{0.8}$S$_2$] – a layered lithium superionic conductor. *Energy Environ. Sci.* **2016**, *9*, 2578–2585.

47. Yang, K.; Dong, J.; Zhang, L.; Li, Y.; Wang, L.; Stevenson, J., Dual Doping: An Effective Method to Enhance the Electrochemical Properties of Li$_{10}$GeP$_2$S$_{12}$-Based Solid Electrolytes. *J. Am. Ceram. Soc.* **2015**, *98*, 3831–3835.

48. Graf, V. H. A; Schäfer, H. Zur Strukturchemie der Alkalisalze der Tetrathiosäuren der Elemente der 5. Hauptgruppe. *Z. Anorg. Allg. Chem.* **1976**, *425*, 67–80.

49. Zhang, L.; Grytsiv, A.; Kerber, M.; Rogl, P.; Bauer, E.; Zehetbauer, M. J.; Wosik, J.; Nauer, G. E. MmFe$_4$Sb$_{12}$- and CoSb$_3$-based nano-skutterudites prepared by ball milling: Kinetics of formation and transport properties. *J. Alloys Compd.* **2009**, *481*, 106–115.

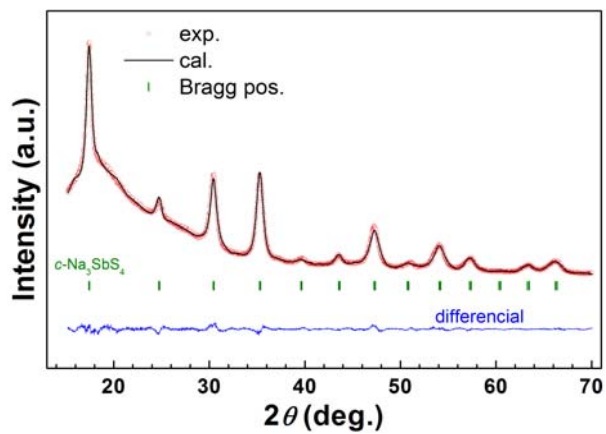

**Figure 1.** Rietveld refinement patterns of high energy ball-milled $Na_3SbS_4$ fitting with cubic models.

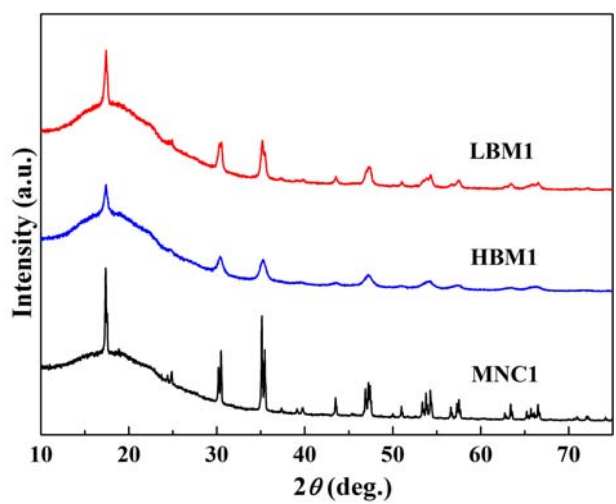

**Figure 2.** XRD profiles of MNC1, HBM1, and LBM1, prepared by the methods of melting, HBM, and LBM, respectively.

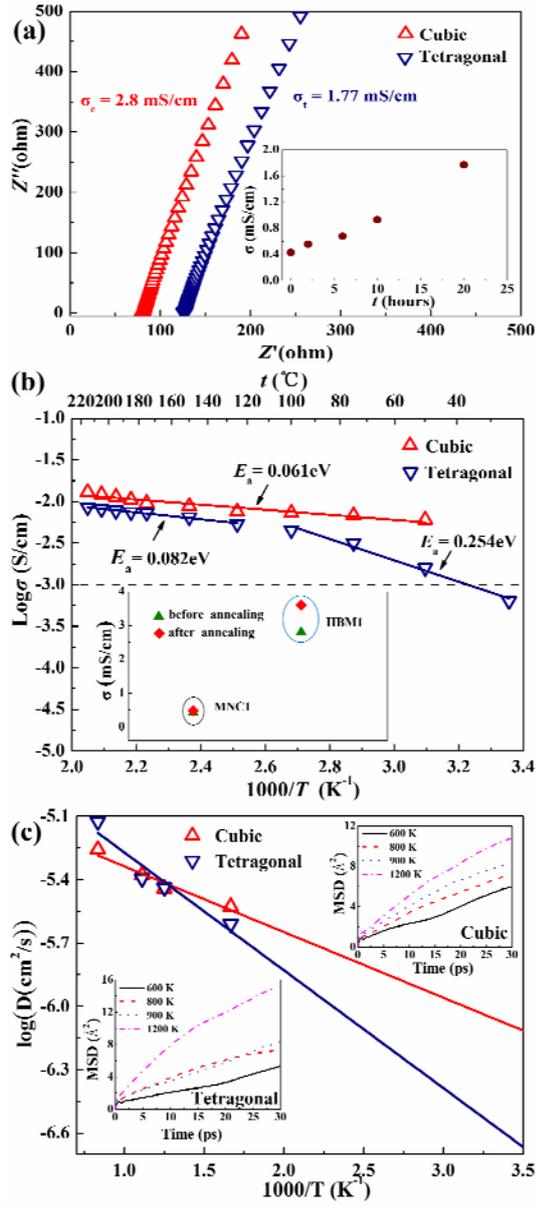

**Figure 3.** (a) Nyquist impedance plot of at room temperature. The inset shows the ionic conductivity of $t$-$Na_3SbS_4$ as a function of LBM time. (b) Arrhenius conductivity plots of HBM1 ($c$-$Na_3SbS_4$) and MNC1 ($t$-$Na_3SbS_4$) in the temperature range of room temperature to 220 °C. The inset shows the ionic conductivity before and after annealing at 120 °C for HBM1 and MNC1. (c) Arrhenius plots of AIMD simulated $Na^+$ diffusivities for $c$- and $t$-$Na_{2.94}SbS_4$ with 2% Na vacancies. The insets display MSD of Na simulated at various temperature. For tetragonal structure Na vacancies were set to locate in $2a$ sites.